\documentclass[a4paper]{jpconf}
\usepackage{graphicx}

\newcommand{\ter}[1]{\ensuremath{\mathrm{#1}}}

\begin{document}
\title{Deep learning techniques for energy clustering in the CMS ECAL}

\author{Davide Valsecchi$^{1,2}$ for the CMS Collaboration}

\address{$^1$ ETH Zurich, $^2$ INFN and Universit\`a degli studi di
  Milano-Bicocca}

\ead{davide.valsecchi@cern.ch}

\begin{abstract}
  The reconstruction of electrons and photons in CMS depends on
  topological clustering of the energy deposited by an incident
  particle in different crystals of the electromagnetic calorimeter
  (ECAL). These clusters are formed by aggregating neighbouring
  crystals according to the expected topology of an electromagnetic
  shower in the ECAL. The presence of upstream material (beampipe,
  tracker and support structures) causes electrons and photons to
  start showering before reaching the calorimeter. This effect,
  combined with the 3.8T CMS magnetic field, leads to energy being
  spread in several clusters around the primary one. It is essential
  to recover the energy contained in these satellite clusters in order
  to achieve the best possible energy resolution for physics
  analyses. Historically satellite clusters have been associated to
  the primary cluster using a purely topological algorithm which does
  not attempt to remove spurious energy deposits from additional
  pileup interactions (PU). The performance of this algorithm is
  expected to degrade during LHC Run 3 (2022+) because of the larger
  average PU levels and the increasing levels of noise due to the
  ageing of the ECAL detector. New methods are being investigated that
  exploit state-of-the-art deep learning architectures like Graph
  Neural Networks (GNN) and self-attention algorithms. These more
  sophisticated models improve the energy collection and are more
  resilient to PU and noise, helping to preserve the electron and
  photon energy resolution achieved during LHC Runs 1 and 2. This work
  will cover the challenges of training the models as well the
  opportunity that this new approach offers to unify the ECAL energy
  measurement with the particle identification steps used in the
  global CMS photon and electron reconstruction.
\end{abstract}

\section{Introduction}
The CMS~\cite{cms} electromagnetic calorimeter (ECAL)~\cite{ecal-tdr}
is a homogeneous calorimeter made of 75848 lead tungstate (PbWO$_4$)
scintillating crystals, located inside the CMS superconducting
solenoid magnet. It is made of a barrel part (EB) covering the region
of pseudorapidity $|\eta| < 1.48$ with 61200 crystals and two endcaps
(EE), which extend the coverage up to $|\eta| < 3.0 $ with 7324
crystals each.  Scintillation light is detected with avalanche
photodiodes (APD) in the barrel and vacuum phototriodes (VPT) in the
endcaps.  The ECAL is crucial for the identification and
reconstruction of photons and electrons, and the measurement of jets
and of missing transverse momentum. The electrons and photons are
typically reconstructed up to $|\eta| < 2.5$, the region covered by
the tracker, while jets are reconstructed up to $|\eta| < 3.0 $.

Several algorithms are stacked on top of each other to reconstruct
electrons and photons candidates from the measurement of scintillation
light in each single crystal in the ECAL
detector~\cite{egamma-paper-run2}.  The first step builds the ECAL
\emph{Rechits}, the measurement of the amount of energy deposited in
each crystal at each LHC bunch crossing (BX). The second step, called
\emph{PFClustering} (Particle Flow Clustering), builds the simplest
form of energy clusters, looking for crystals with local maxima of
energy (\emph{seeds}) and associating the neighbor crystals to them.

A single electron or photon usually leaves more than one cluster of
energy in the ECAL detector. The electron, bending in the strong
magnetic field of the CMS solenoid (3.8 T) while passing through the
Pixel and Tracker detectors, emits bremsstrahlung photons that will
leave a trace of small energy clusters in the ECAL detector near the
main impact point, mostly extended in the transverse $R-\phi$
plane. The photon, instead, is converted to electron-positron pairs
interacting with the several layers of the inner detectors of CMS,
thus also depositing multiple clusters of energy in ECAL.  An
additional clustering algorithm, called \emph{SuperClustering} is
therefore needed for the electron and photon reconstruction in order
to improve the energy resolution by taking into account the energy of
the secondary clusters.

The SuperCluster (SC) candidate is built using only ECAL local
information, therefore it is the object used for the calibration of
the ECAL detector response. The SuperCluster is also one of the inputs
of the Particle Flow (PF) CMS global event
reconstruction~\cite{CMS-PRF-14-001} which combines optimally the
information from all the sub-detectors.  Electrons are identified as a
primary charged-particle track linked to ECAL SuperClusters, whereas
photons are identified as ECAL deposits not linked to any extrapolated
track.  A comprehensive description of electron and photon
reconstruction during Run II can be found in
Ref.~\cite{egamma-paper-run2}.


\section{SuperClustering algorithms}

\begin{figure}[h]
  \centering
  \includegraphics[width=0.5\textwidth]{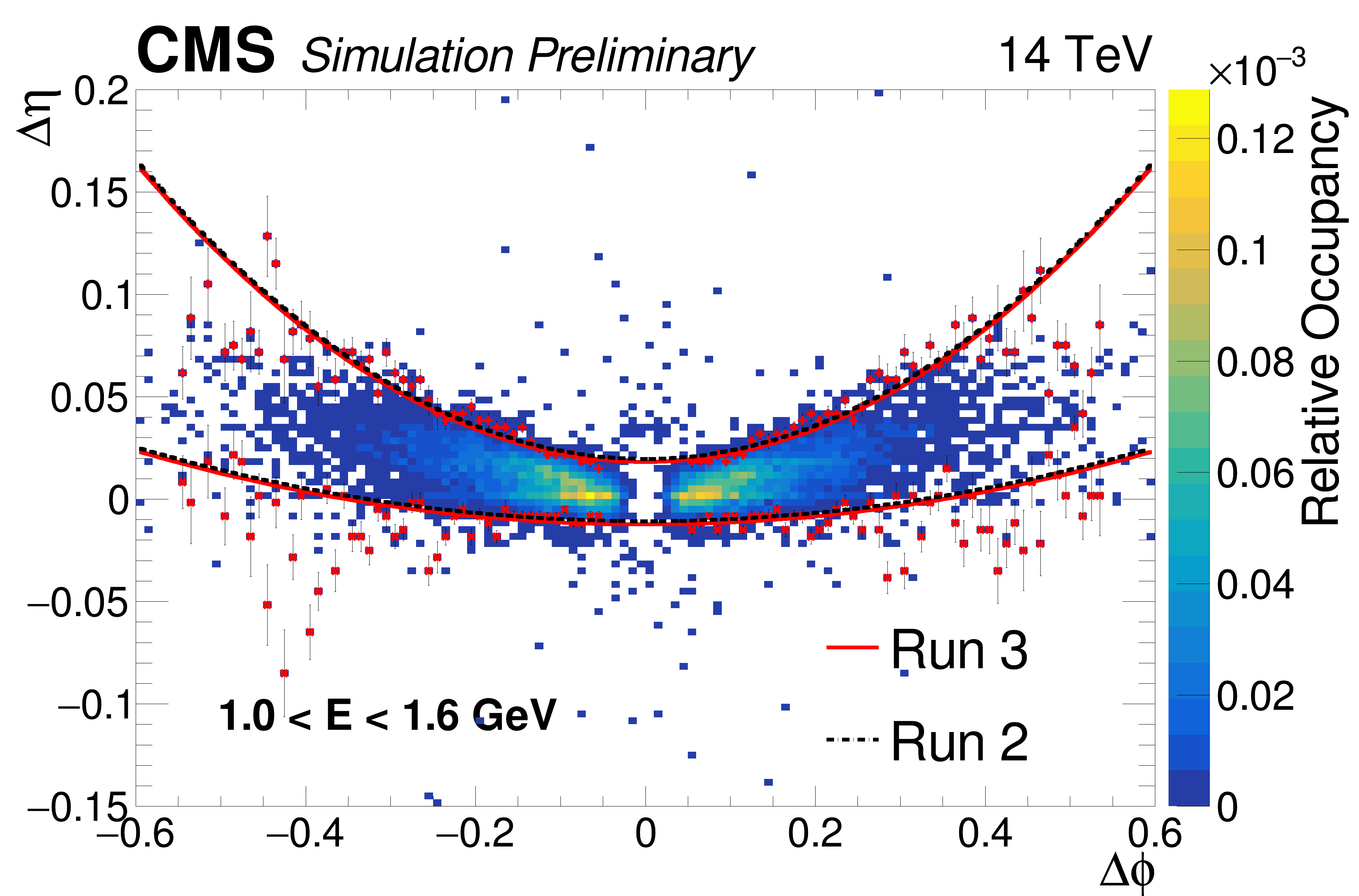}
  \caption{\label{fig:mustache} Shape of the SuperCluster as selected
    by the Mustache algorithm, obtained by accumulating many events in
    the bin $\eta\in[0.5,0.8]$, $E_T\in[1.0,1.6]$~GeV of the seed.  98\% of the
    electromagnetic energy is contained in the Mustache area.}
\end{figure}

The SuperClustering algorithm in place during the LHC Run II (in Run I
other algos were used) is based on purely geometrical criteria and
called \emph{Mustache}: in fact the magnetic field causes the low
$p_T$ constituents of electromagnetic showers to slightly bend in
$\eta$ as they follow a helical path, giving the showers a
characteristic mustache shape in the $\Delta\eta - \Delta\phi$ plane
(see Fig.~\ref{fig:mustache}).  A nearby cluster is associated to the
seed if it falls inside the $\Delta\eta -\Delta\phi$ region delimited
by two parabolas parametrized by the $\eta$ position of the seed and
the energy of the cluster, and by a dynamic $\Delta\phi$ interval
which depends instead only on the transverse energy of the
cluster. The parameters of this spatial selection are optimized to
contain 98\% of the electromagnetic (EM) energy of the shower in
several bins of energy of the seed and position along the detector.
Fig.~\ref{fig:mustache} shows an example of the Mustache shape.

Being the Mustache SuperClustering a purely geometrical algorithm, it
has a high signal efficiency, but it does not attempt to identify and
remove clusters generated from electronics noise or from additional
low energy p-p interactions overlapping in the event (pileup or PU).
The increase of the pileup level and the expected increase of noise
contamination during Run III of LHC due to the ECAL detector ageing,
makes the development of a more performant SuperClustering algorithm
based on supervised machine learning (ML) methods desirable.

The ML model architecture chosen to tackle the SuperClustering problem
involves Graph Convolutional Networks
(GCN)~\cite{Shlomi:2020gdn,gnn,Battaglia:2016jem,xin2020graph} and
Self-Attention (SA)
layers~\cite{vaswani2017attention,Mikuni:2021pou}. This class of
models permits working on a set of related objects inferring
properties about the single elements, or the overall set. One of the
advantages over other architectures is that the number of objects to
analyze can be different for each event. This type of architecture
have also been recently explored successfully for developing a ML
approximation of the full CMS PF algorithm\cite{Pata:2021oez}.

The novel \emph{DeepSuperCluster} (DeepSC) model is designed to
perform three tasks at the same time: (i) SuperClustering: identify
which cluster should be associated to the seed cluster to build the
optimal SuperCluster; (ii) Energy regression: extract the correction
factor needed to restore the generator-level particle
energy. Currently this regression is trained separately on the
Mustache SuperClusters to obtain the final energy estimation for ECAL
calibration; (iii) Identification: classify the different types of
energy deposits to discriminate between isolated electron/photon, or
particles from jets. This note describes the performance of a model
optimized to perform the clustering task, while the energy regression
and object identification ones will be optimized in a second step.

\section{DeepSC training sample and model architecture}

\subsection{Training sample}
The model is trained on a sample of $2\cdot 10^6$ photons and
electrons, generated uniformly in $\eta$ and $p_{T} =[1, 100]$ GeV,
with the full Geant4-based CMS Monte-Carlo simulation at 14
TeV. Energy deposits from noise and pileup interactions are
superimposed on the signal. A pileup scenario with the number of true
interactions uniformly distributed between [55,75] is used.

An optimal true association between the seeds and the clusters in the
event is built by tracking the EM shower produced by original
generator-level electron or photon in the Geant4 simulation, and
storing the information about the amount of energy deposited in each
ECAL crystal.  This information allows to apply a precise matching of
each cluster to the generator-level particle with energy thresholds
optimized to obtain the best possible SuperCluster energy resolution,
and taking fully into account effects from the energy deposition and
reconstruction in the calorimeter.

The model is applied on regions of the ECAL detector around each
energetic cluster ($E_T > 1$ GeV), called seed, using the following
information: (i) Energy, position and number of crystals of each
cluster
; (ii) Information relative to the seed for each cluster:
$\Delta\eta(\ter{cl},\ter{seed})$, $\Delta\phi(\ter{cl},\ter{seed})$,
$\Delta E_T(\ter{cl},\ter{seed})$, $\Delta En(\ter{cl},\ter{seed})$;
(ii) List of $N$ rechit information for each $i$-th cluster $[(i\eta
  ,i \phi, iz , En)_0, \cdots, (i\eta ,i \phi, iz , En)_{N}]^i $;
(iiii) Summary information for each region of the detector: maximum,
minimum, and average of the cluster related features.

\subsection{Model architecture}
\begin{figure}[h]
  \centering
  \includegraphics[width=0.92\textwidth]{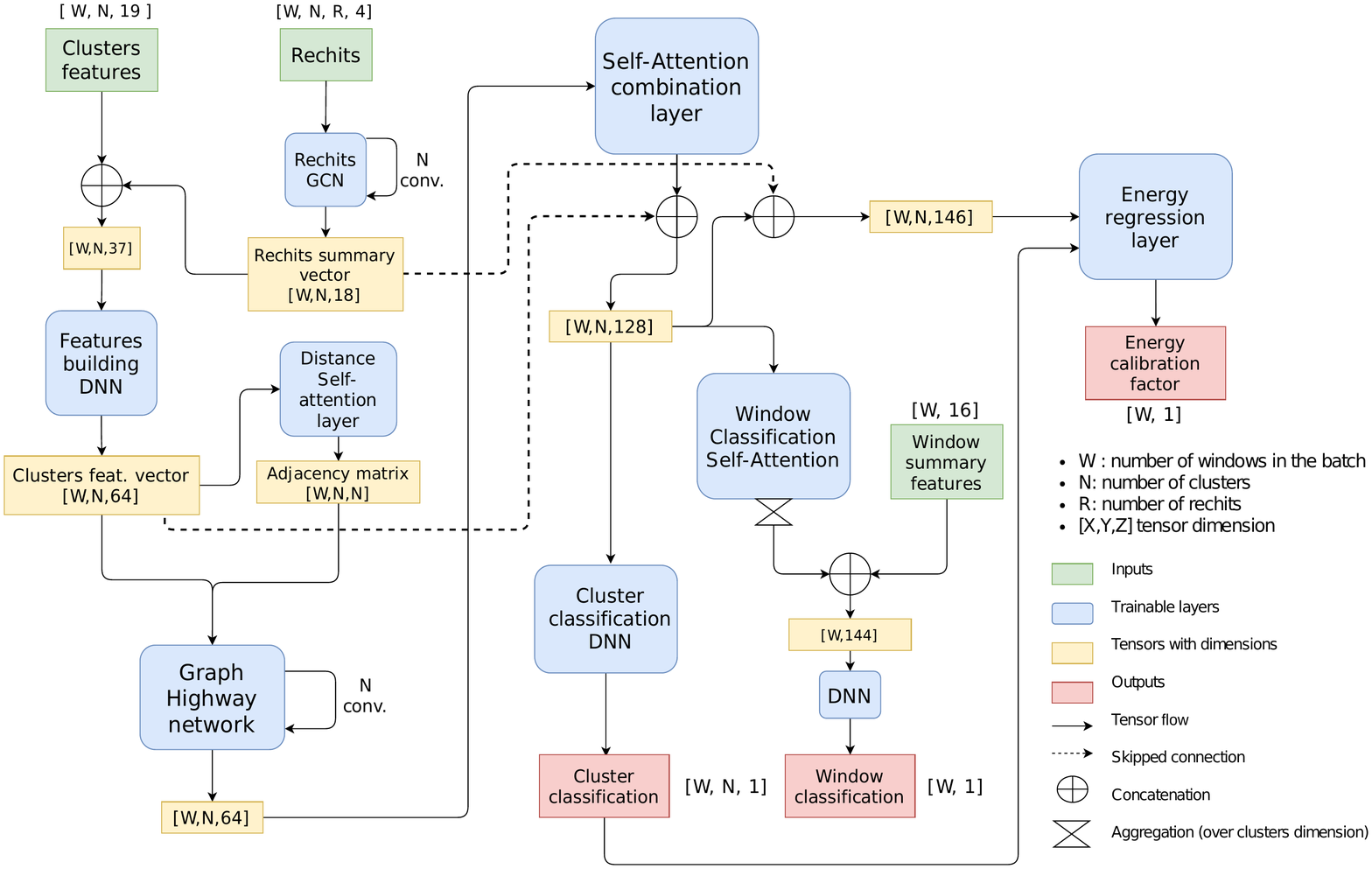}
  \caption{\label{fig:deepsc_model} Schema of the DeepSC model
    depicting the models inputs, layers and tensors flowing through.}
\end{figure}

Fig.~\ref{fig:deepsc_model} represents the architecture of the model
and the dimensions of tensors processed with it. The model is
conceptually organized in four steps: encoding, graph building, graph
elaboration, decoding (for each output).  The model has been fully
implemented in TensorFlow library version
2.3~\cite{tensorflow_developers_2021_5593257}.

\begin{description}
\item[Encoding] A GNN layer is applied on the rechits of each cluster
  to obtain a single fixed-size vector representing summary rechit
  information for each cluster.  Then, a simple feature-extracting DNN
  is applied on the input vector of each cluster and rechit latent
  vector to obtain a 64 features vector for each cluster.
\item[Graph building] The adjacency matrix $A_{ij}$ defining the
  distance between each cluster is defined
  dynamically~\cite{Shlomi:2020gdn} assigning to each cluster a
  3D-space coordinates vector with a SA layer and computing the
  Euclidean distance. Therefore, the relative importance of the
  interaction between the clusters, i.e. their distance in the graph,
  is learned automatically during the training.
\item[Graph elaboration] A graph convolutional layer called Graph
  Highway Network~\cite{xin2020graph} is applied on the clusters graph
  and then a SA layer is a applied to extract a features vector for
  each cluster. The application of the graph network implies a passage
  of information between the clusters.
\item[Decoding] Output information is extracted from the latent
  space. First, a simple DNN network is applied on each cluster latent
  vector to get the classification output. Then, all the clusters
  latent vectors are combined with a different SA layer and then
  aggregated in a single vector to obtain the detector region
  classification. An energy regression layer can be applied in the
  same fashion.
\end{description}

\section{Performance comparison}

\begin{figure}[h]
  \begin{minipage}{0.48\textwidth}
    \includegraphics[width=\textwidth]{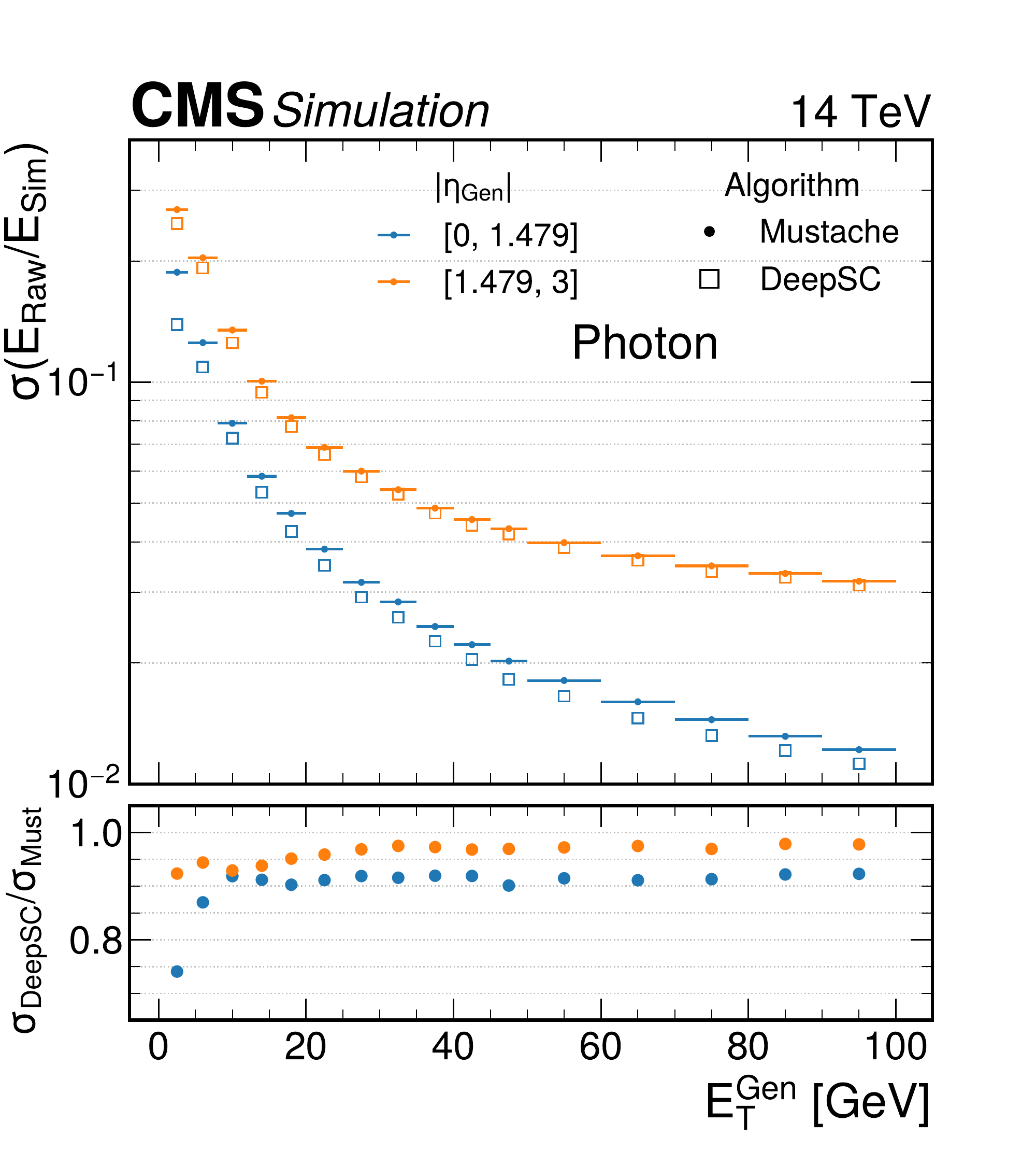}
    \caption{\label{fig:res-en}Comparison between the DeepSC and
      Mustache energy resolution in bins of generator-level particle
      $E_T^{\ter{Gen}}$.}
  \end{minipage}
  \hspace{2em}
  \begin{minipage}{0.48\textwidth}
    \includegraphics[width=\textwidth]{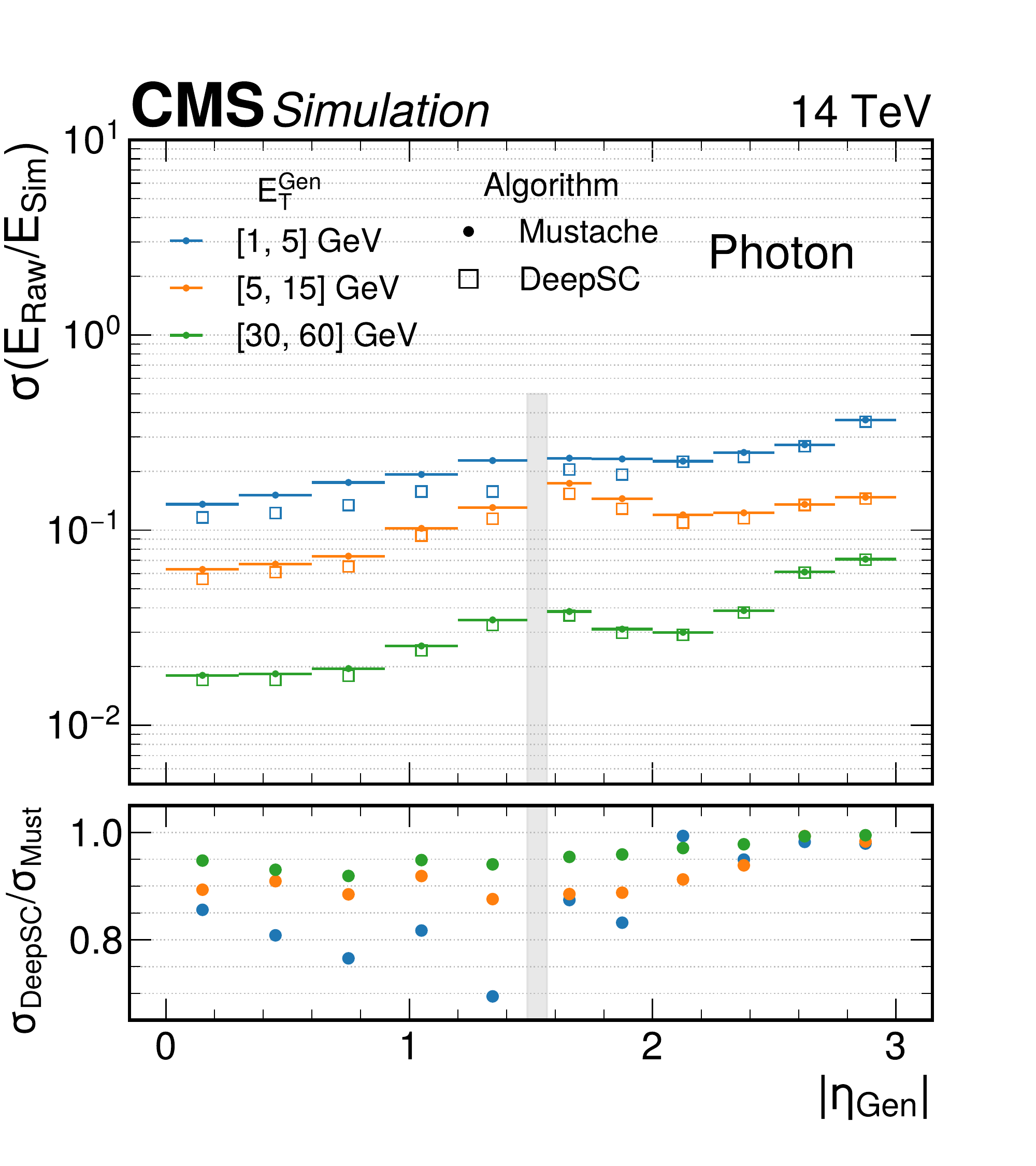}
    \caption{\label{fig:res-eta}Comparison between the DeepSC and
      Mustache energy resolution in bins of generator-level particle
      position $|\eta|$.}
  \end{minipage}
  \begin{center}
    \begin{minipage}{0.48\textwidth}
      \includegraphics[width=\textwidth]{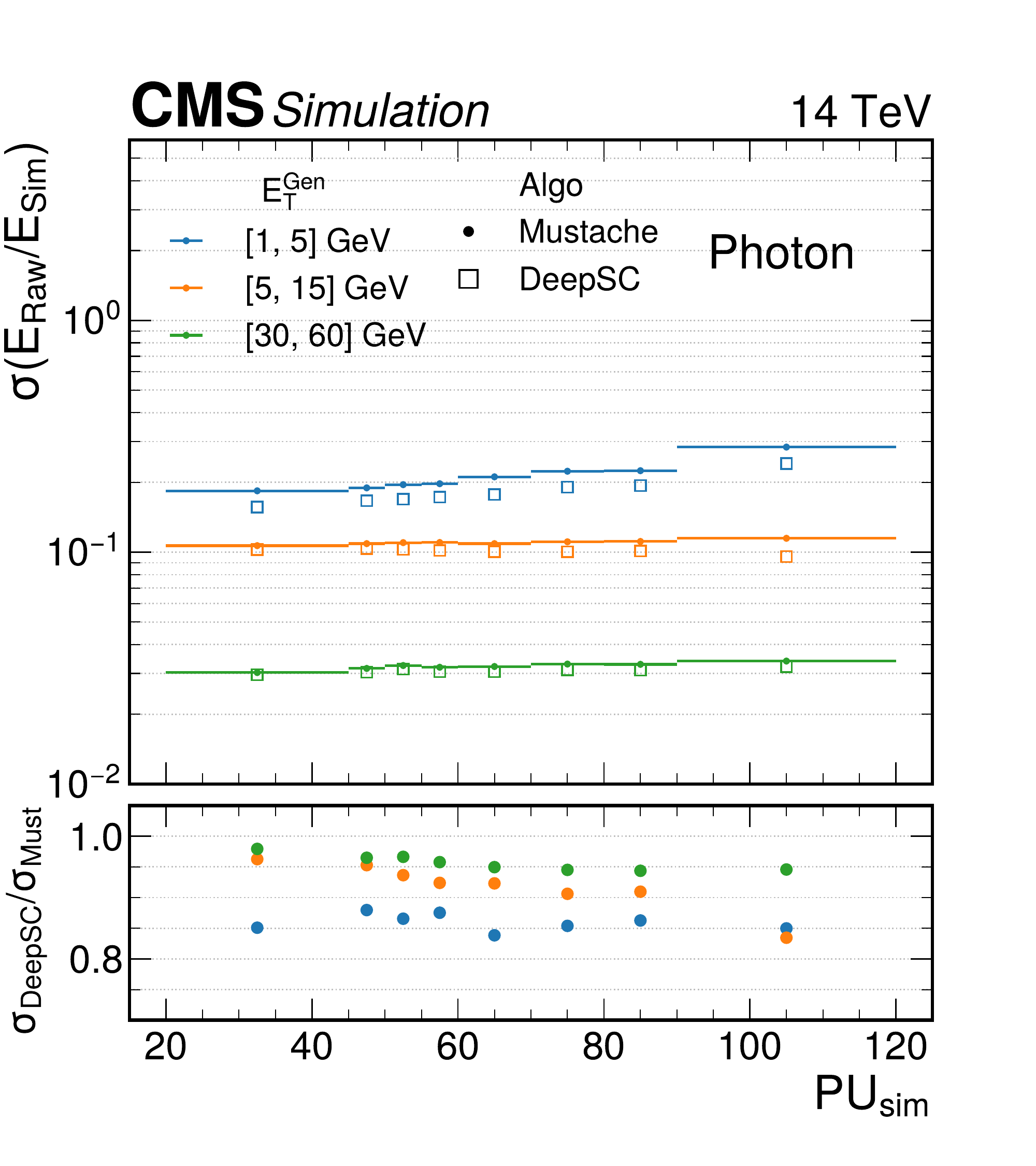}
      \caption{\label{fig:res-pu}Comparison between the DeepSC and
        Mustache energy resolution in bins of number of simulated PU
        vertices.}
    \end{minipage}
  \end{center}  
\end{figure}

The performance of the DeepSC and Mustache algorithms is
compared~\cite{CMS-DP-2021-032} in terms of the resolution of the
uncorrected reconstructed SuperCluster energy $E_{\ter{Raw}}$ divided
by the true simulated energy deposits in ECAL $E_{\ter{Sim}}$.  The
resolution is analyzed for electrons and photons versus the
generator-level particle $E_T$ (Fig.~\ref{fig:res-en}), position
$|\eta|$ in the detector (Fig.~\ref{fig:res-eta}), and for different
number of simulated PU vertices in the event (Fig.~\ref{fig:res-pu}).
The resolution is computed as half of the difference between the 84\%
quantile and the 16\% quantile of the $E_{\ter{Raw}}/E_{\ter{Sim}}$
distribution in each bin.

The DeepSC algorithms keeps the signal efficiency similar to the
Mustache one, but largely reduces the noise and PU contamination. The
DeepSC largely improves the energy resolution at low energy and in the
$|\eta| \in [1,2]$ region, where there are more secondary emissions at
lower energy from electrons and photons, due to the higher material
budget in front of ECAL. The DeepSC algorithm also makes the
dependence on the energy resolution versus the number of PU vertices
flatter, without using any explicit input information related to PU.

\section{Conclusions}
The DeepSC algorithm is a promising ML based alternative to the CMS
ECAL Mustache algorithm. It is more robust against pileup and it
outperforms the Mustache in terms of energy purity and capability of
removing spurious energy deposits from noise and pileup, hence
improving the uncorrected energy resolution of the ECAL detector.

\section*{References}

\bibliography{biblio}

\end{document}